\documentclass[twocolumn,aps,floatfix,prc]{revtex4-1}
\usepackage{graphicx}
\usepackage{dcolumn}
\usepackage{bm}
\usepackage{rotating}
\usepackage{longtable}
\usepackage{float}
\usepackage{eucal}
\usepackage{csquotes}
\usepackage{xcolor}
\usepackage{amsmath}
\usepackage{amssymb}
\usepackage{natbib}
\usepackage[T1]{fontenc}
\usepackage[pagewise]{lineno}
\usepackage{amstext}
\begin{document}
\title{Quasi-${\bm\gamma}$ band in $^{\text{114}}$Te}
\author{Prithwijita Ray}
\author{H. Pai}
\email{h.pai@saha.ac.in; hari.vecc@gmail.com}
\author{Sajad Ali}
\author{Anjali Mukherjee}
\author{A. Goswami\footnote{Deceased}}
\affiliation{Saha Institute of Nuclear Physics, 1/AF Bidhan Nagar, Kolkata 700064, India}
\affiliation{Homi Bhabha National Institute, Training School Complex, Anushaktinagar, Mumbai 400094, India}
\author{S. Rajbanshi}
\affiliation{Department of Physics, Presidency University, Kolkata 700073, India}
\author{Soumik Bhattacharya}
\author{R. Banik}
\author{S. Nandi}
\author{S. Bhattacharyya}
\author{G. Mukherjee}
\author{C. Bhattacharya}
\affiliation{Variable Energy Cyclotron Centre, 1/AF Bidhan Nagar, Kolkata 700064, India }
\affiliation{Homi Bhabha National Institute, Training School Complex, Anushaktinagar, Mumbai 400094, India}
\author{S. Chakraborty}
\affiliation{Nuclear Physics Group, Inter-University Accelerator Centre, New Delhi 110067, India.}
\author{G. Gangopadhyay}
\affiliation{Department of Physics, University of Calcutta, Kolkata 700009, India}
\author{Md. S. R. Laskar}
\author{R. Palit}
\affiliation{Department of Nuclear and Atomic Physics, Tata Institute of Fundamental Research, Mumbai 400005, India}
\author{G. H. Bhat, S. Jehangir}
\affiliation{Department of Physics, Sri Pratap College, Srinagar 190001, India.}
\affiliation{Cluster University Srinagar, Jammu and Kashmir, 190001, India.}
\author{J. A. Sheikh}
\affiliation{Cluster University Srinagar, Jammu and Kashmir, 190001, India}
\affiliation{Department of Physics, University of Kashmir, Srinagar 190006, India.}
\author{A. K. Sinha}
\affiliation{UGC-DAE Consortium for Scientific Research, University Campus, Khandwa Road, Indore 452017, India}
\author{S. Samanta}
\author{S. Das}
\author{S. Chatterjee}
\author{R. Raut}
\author{S. S. Ghugre}
\affiliation{UGC-DAE Consortium for Scientific Research, Kolkata centre, Kolkata 700098, India}
\date{\today}
\begin{abstract}
The low-lying non-yrast states in $^{114}$Te have been investigated using the Indian National Gamma Array through the fusion-evaporation reaction $^{112}$Sn($^{4}$He, 2n) at a beam energy of 37 MeV. Eight new $\gamma$-transitions have been placed in the level scheme to establish the quasi-$\gamma$ band in this nucleus. Spin and parity of several excited states have been assigned from the present spectroscopy measurements. The comparison of experimental results on the observed bands with the Interacting Boson Model (IBM) and Triaxial Projected Shell Model (TPSM) confirming the existence of the quasi-$\gamma$ band structure in the $^{114}$Te nucleus.
\end{abstract}
\maketitle
\section{Introduction}
A large number of collective and non-collective states have been observed at lower excitation energies exhibiting a shape coexistence as being mainly caused by proton 2p-2h excitation along with the spherical ground states across the \textit{Z = 50} shell closure. For deformed nuclei, the nuclear shapes have traditionally been described in terms of $\beta$ and $\gamma$ parameters, where the former specifies the ellipsoidal quadrupole deformation and the latter the degree of axial asymmetry. The most common low-lying collective excitation, namely $\gamma$-vibration has been extensively reported in several mass regions throughout the nuclear chart over decades~\cite{bohr,lw,rk,kusakari}. $\gamma$-bands are associated with ellipsoidal oscillation of nuclear shape. This kind of phenomenon is favourable when the potential energy surfaces are found to be soft for both $\beta$ and $\gamma$ deformation parameters, ensuring that shape polarization can take place.

In the last decade, several groups have tried to tackle the question of whether the pure vibrational structure can or cannot be observed in the nuclei in the $ Z \approx 50 $ region. In this context, the work by Garret and Wood is useful to understand the interplay between $ \gamma $-soft and vibrational character of the low-lying collective excitations in these nuclei~\cite{garret2010}. Consequently, the quasi-$ \gamma $ bands and co-existence of different nuclear shapes associated with intruder configuration at low angular momentum were observed frequently in these nuclei~\cite{garret2012,garret2014,garret2016,Petrache,Spieker}. Recently, a $ \gamma $-band built on the shape-coexisting intruder configuration is reported in Cd isotopes~\cite{garret2019}. Well-developed quasi-$\gamma$ bands, built on the 2$^+_2$ state, were reported in even-even Xe and Ba isotopes, having both favoured and unfavoured partners~\cite{kusakari,saikat}. The $ ^{120,122,124} $Te nuclei were reported to be soft-triaxial in nature~\cite{ms} and hence, the occurrence of quasi-$ \gamma $ bands is also expected in these nuclei. Consequently, the quasi-$ \gamma $ bands have been identified recently in heavier mass $ ^{124,126} $Te isotopes, but, without any in-band $ \gamma $ transitions \cite{shashi}. However, the quasi-$ \gamma $ bands have not been studied well in any of the lighter mass Te isotopes. Only the favoured partner of the quasi-$\gamma$ band was reported tentatively in $ ^{118} $Te~\cite{118te}. As the excitation energy of the quasi-$ \gamma $ vibrational states is sensitive to a different kind of critical point symmetry \cite{Casten,McCutchan,Casten-1}, therefore, it is important to search for the quasi-$ \gamma $ bands in Te isotopes systematically to investigate the shape evolution of Te nuclei. With this motivation, an attempt has been made to search for the quasi-$ \gamma $ band in $ ^{114} $Te.

The low-lying non-yrast structure of neutron deficient $ ^{114} $Te has not been studied extensively. Only a few experimental studies have been reported~\cite{114te1, 114te2, 114te3}. This nucleus is shown to mimic the U(5) dynamical symmetry based on the energy ratio, R$_{4/2}$ of 2.09. However, the transition probabilities [B(E2)], extracted from the lifetime measurement, are found in strong contradiction with the same estimated by theoretical interacting boson model (IBM) within the U(5) limit~\cite{Mol,Cakirli}. In this context, it is worth noting that, a reasonable description of the yrast B(E2) values was reported recently from a large-scale-shell-model study~\cite{qi}. In order to elucidate the underlying structure of $^{114}$Te, an experiment was carried out using $\alpha$-beam, delivered by the K-130 cyclotron of Variable Energy Cyclotron Centre, Kolkata.

\section{EXPERIMENTAL DETAILS}

The excited states in $^{114}$Te were populated through the fusion-evaporation reaction $^{112}$Sn($^{4}$He, 2n)$^{114}$Te at a beam energy of 37 MeV delivered by the K-130 cyclotron accelerator of the Variable Energy Cyclotron Centre, Kolkata. A self-supporting isotopically-enriched (99.6\%) $^{112}$Sn foil of effective thickness 4.5 mg/cm$^2$ was used as a target~\cite{112Sn}. The de-excited $\gamma$-rays were detected by the Indian National Gamma Array (INGA), stationed at the Variable Energy Cyclotron Centre, Kolkata comprising of seven Compton-suppressed HPGe clover detectors and one LEPS (Low Energy Photon Spectrometer) detector~\cite{veccinga}. These detectors were arranged at different angles ($\theta$) with respect to the beam direction; four clover detectors at $\theta$ = 90$^\circ$ (2 in-plane and 2 out-of-plane), two clover detectors at $\theta$ = 125$^\circ$, one clover detector at $\theta$ = 40$^\circ$ and the LEPS detector at $\theta$ = 40$^\circ$. The energy and efficiency calibrations have been performed using the $^{133}$Ba and $^{152}$Eu radioactive 
sources placed at the target position of the INGA set up. 

The pulse processing and data acquisition system was based on PIXIE-16 12-bit 250 MHz digitizer modules manufactured by 
XIA LLC and running on firmware conceptualized by UGC-DAE CSR, Kolkata Centre~\cite{rr1}. 
Time stamped list mode data was acquired with event trigger generated from coincidence firing of at least two Compton suppressed 
clover detectors. Around 1.5 $\times$ 10$^8$ such events were acquired during the experiment.
The data was processed into spectra, $\gamma \gamma$ matrices and $\gamma \gamma \gamma$ cube
using the IUCPIX~\cite{rr1,rr2} package, developed at UGC-DAE CSR, Kolkata Centre, and 
analysed using RADWARE~\cite{rw} and INGASORT~\cite{rb} packages. 

\section{DATA ANALYSIS PROCEDURES}

The spin of an excited nuclear state can be determined from the multipolarity of the $\gamma$-ray. The multipolarities of the $\gamma$-ray transitions were determined from the angular correlation analysis using the method of directional correlation from the oriented states (DCO) ratio~\cite{bkm}. For the DCO ratio analysis, an asymmetric matrix was constructed using the coincidence events registered in the detectors placed at angles 125$^{\circ}$ and 90$^{\circ}$ with respect to the beam axis. The DCO ratio of the $\gamma$-ray transition ($\gamma_1$) at angle $\theta_1$ = 125$^{\circ}$ gated by the transition of known multipolarity ($\gamma_2$) at angle $\theta_2$ = 90$^{\circ}$ is defined as,

\begin{equation}
\label{rdco}
R_{DCO} = \frac{I_{\gamma_1} ~ at~ \theta_1, ~gated ~by ~\gamma_2 ~at ~\theta_2}
               {I_{\gamma_1} ~at ~\theta_2, ~gated ~by ~\gamma_2 ~at ~\theta_1}
\end{equation}

For the same multipolarity of the stretched transitions $\gamma_1$ and $\gamma_2$, the value of R$_{DCO}$ is close to unity. The value of R$_{DCO}$ for the transitions $\gamma_1$ and $\gamma_2$ with different multipolarities depends on the detector angles ($\theta_1$ and $\theta_2$) and the mixing ratios ($\delta$) associated with the transitions. The validity of the measured R$_{DCO}$ values were checked with the known transitions in $^{114}$Te~\cite{114te3} and have been compared with the theoretical values calculated using the code {\scriptsize{ANGCOR}}~\cite{angcor}. In the present geometry, the calculated value of R$_{DCO}$~\cite{angcor} for a pure dipole(quadrupole) transition gated by a stretched quadrupole(dipole) transition is 0.70(1.50). Measured experimental values for the pure dipole (936.2 keV, 7$^{-}$ $\rightarrow$ 6$^{+}$, $E1$) and quadrupole (774.8 keV, 4$^{+}$ $\rightarrow$ 2$^{+}$, $E2$) transitions of $^{114}$Te are, 0.70 (0.08) and 1.60 (0.18), respectively, in good agreement with the calculated values ~\cite{angcor}. The width of the sub-state population (${\sigma}/{J}$) required for the R$_{DCO}$ calculation is assumed to be of 0.37 for the present experiment.


\begin{figure}[!]
\begin{center}
\includegraphics[width=\columnwidth]{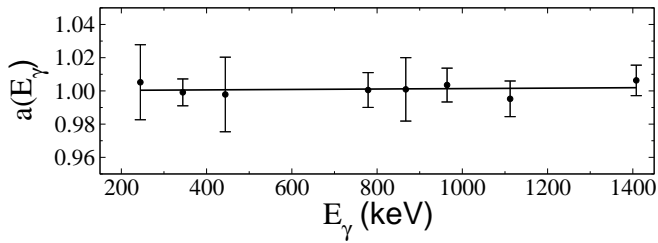}
\caption{The asymmetry correction factor a(E$_\gamma$) at different gamma ray energies from $^{152}$Eu source. The solid line corresponds to a linear fit of the data.}
\label{gas}
\end{center}
\end{figure}


\begin{figure}[!]
\begin{center}
\includegraphics*[width=\columnwidth]{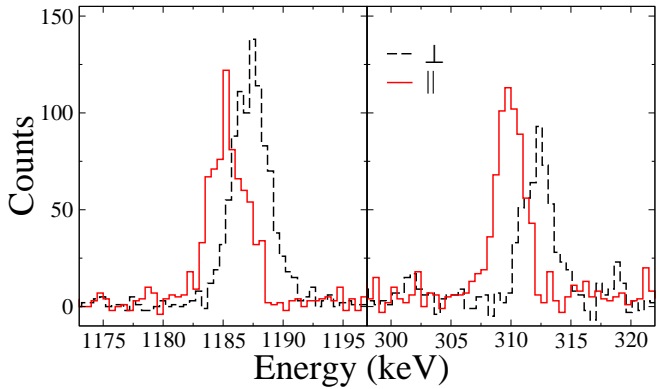}
\caption{The perpendicular (black dashed line) and parallel (red solid line) components of 
the two $\gamma$ rays in $^{114}$Sb, obtained from the linear polarization analysis in the present work. The 309.2 keV transition (right) is known as magnetic type transition whereas the 1184.9 keV transition (left) is an electric type transition. The perpendicular component was shifted in energy for clarity.}
\label{pol}
\end{center}
\end{figure}

Definite parities of the excited states have been designated from the linear polarization asymmetry ratio ($ \Delta_{\text{asym}} $), extracted from the parallel and perpendicular scattering of the $\gamma$ photons inside the detector medium~\cite{bp1,bp2,hp}. The measured value of $\Delta_{\text{asym}}$ provides a qualitative idea about the electric or magnetic nature of the transitions. The $\Delta_{\text{asym}}$ value for a $\gamma$-transition has been deduced using the relation:

\begin{equation}
\label{ipdco}
\Delta_{\text{asym}} = \frac{a(E_\gamma) N_\perp - N_\parallel}{a(E_\gamma)N_\perp + N_\parallel},
\end{equation}
where, $N_\parallel$ and $N_\perp$ are the counts for the actual Compton-scattered $\gamma$ rays in the planes parallel and perpendicular to the reaction plane, respectively. Correction due to the asymmetry in the array and the response of the clover segments, defined by $a(E_\gamma$) = $\frac{N_\parallel} {N_\perp}$, was estimated using unpolarised $^{152}$Eu source. The energy dependence of the parameter $a(E_\gamma$) is obtained using the expression $a(E_\gamma$) = $a$ + $b(E_\gamma$). The fitting, shown in Fig.~\ref{gas}, gives the values of the constants as $a = 1.000(3)$ and $b(E_\gamma$) $\sim$ 10$^{-6}$. A positive value of $\Delta_{\text{asym}}$ indicates an electric ($E$) type transition, whereas, a negative value indicates a magnetic ($M$) type transition. The low energy cut-off for the polarization measurement in the present work was about 200 keV. The validity of this method has been confirmed from the known transitions of energy 1184.9-keV and 309.2-keV in $^{114}$Sb (Fig.~\ref{pol}) which was also produced in the same experiment~\cite{114sb,NDS-114Te}.

\section{EXPERIMENTAL RESULTS}
\begin{figure}[t]
\begin{center}
\includegraphics[height=\columnwidth, angle = -90]{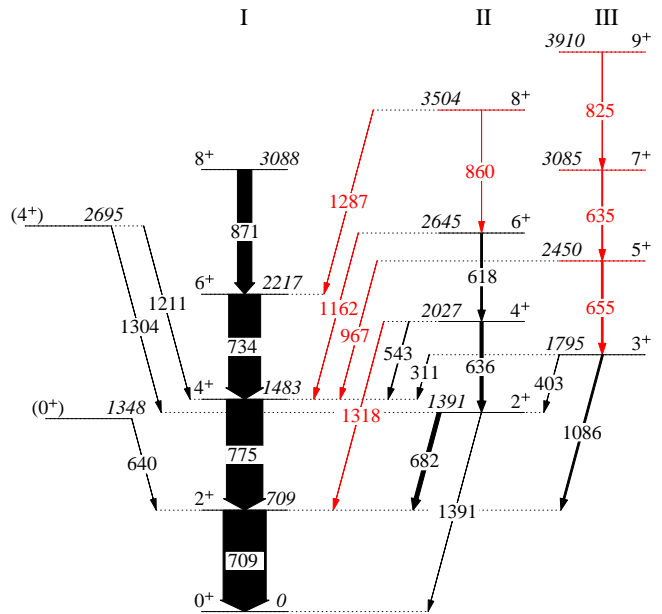}
\caption{Partial level scheme of $^{114}$Te, deduced from this work. Newly identified $\gamma$-rays/levels are marked in red (gray) colour.}
\label{LS}
\end{center}
\end{figure}

\begin{figure*}
\begin{center}
\includegraphics[width=0.95\textwidth]{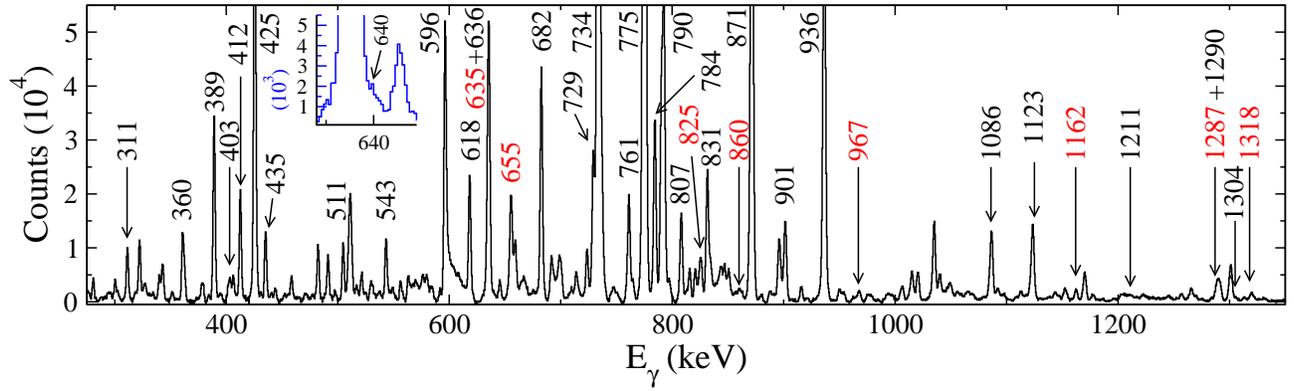}
\end{center}
\caption{Spectrum to show the $ \gamma $ rays observed in coincidence with the 708.7 keV $ \gamma $ transition in $^{114}$Te. Newly observed $ \gamma $ rays are shown in red (gray) color. $ \gamma $ rays, those are marked here but not included in \figurename~\ref{LS}, are already reported in Ref.~\cite{114te2}.}
\label{gate-709}
\end{figure*}

\begin{figure}[h]
\begin{center}
\includegraphics[width=\columnwidth]{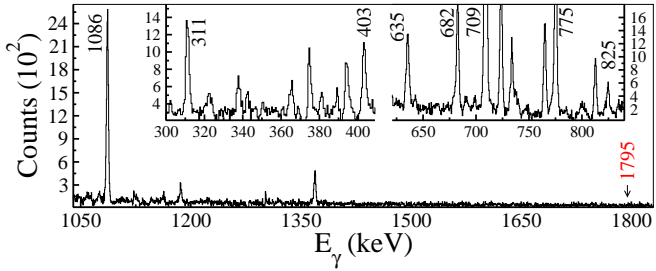}
\caption{Prompt $ \gamma \gamma $ coincidence spectra gated by 655.3-keV transition. The 1795 keV (red marked) transition was reported in Refs.~\cite{zimmerman,NDS-114Te}, but not observed in the present work.}
\label{gate-655}
\end{center}
\end{figure}

Low-lying level scheme of $^{114}$Te (up to $E_{x} \approx 4$ MeV) has been updated in the present work by placing eight new $\gamma$-transitions as shown in Fig.~\ref{LS}. Energy levels were grouped into three bands I - III. The experimental results obtained in the present work were summarized in Table~\ref{table1}. Measured relative intensities of the $\gamma$ rays from the single-gated spectra were normalized for the 708.7-keV transition. Relevant energy gated $\gamma \gamma$ spectra in support of the present placements were presented in Figs.~\ref{gate-709}-\ref{gate-640_1211_1304}. These spectra show the previously reported $\gamma$-rays in $^{114}$Te ~\cite{114te1, 114te2, 114te3} along with the newly observed $\gamma$-rays.

\begin{table*} [ht]
\caption{Initial states ($E_i$) and Energies of $\gamma$-rays ($E_\gamma$), Relative Intensity ($I_\gamma$), DCO ratios (R$_{\text{DCO}}$), linear polarization asymmetry ($\Delta_{\text{asym}}$), mixing ratio ({$\delta$}) and Deduced multipolarity ($ E \lambda / M \lambda $) of the $\gamma$-transitions in \text{$^{114}$Te}.}
\label{table1}
\begin{ruledtabular}
\begin{tabular}{ccccccccc}
$E_i$ (keV) & \text{$E_\gamma$} (keV)\footnote{Uncertainty in $ \gamma $-ray energy is $\pm$ (0.3-0.5) keV.}&
\text{$I_\gamma$}\footnote{Intensities of $\gamma$-rays are normalized to 708.7 keV transition, with $I_{\gamma}$ = 100. Intensity uncertainties does include the errors due the uncertainty in efficiency correction.} &
\textbf{$J_{i}$$^{\pi}$} & \text{$J_{f}$$^{\pi}$} & {$\text R_{\text{DCO}}$} & $\Delta_{\text{asym}}$ & {$\delta$} & $ E \lambda / M \lambda $ \\
\hline
708.7 &708.7& 100.0(5.2) & 2$^{+}$   & 0$^{+}$    &  1.00(0.08)$^f$   & $+$0.12(0.01)  & & \textit{E}2\\

1348.2 &639.5& 0.33(0.05) & (0)$^{+}$\footnote{From Ref.~\cite{NDS-114Te}.}   & 2$^{+}$ &0.94(0.45)$^d$      &   & &(\textit{E}2)\\

1391.2 &682.3& 9.28(0.94) & 2$^{+}$   & 2$^{+}$    & 0.96(0.11)$^d$     & $-$0.07        (0.03)  &2.5& \textit{M}1+\textit{E}2\\

&1391.2& 0.12(0.03) & 2$^{+}$   & 0$^{+}$    &      &   & &\\

1483.5 &774.8& 84.67(8.64) & 4$^{+}$   & 2$^{+}$    & 0.98(0.11)$^d$     & $+$0.13(0.01)  &  & \textit{E}2\\

1794.7 &310.9& 1.00(0.11) & 3$^{+}$   & 4$^{+}$    & 0.63(0.08)\footnote{DCO ratios are obtained from 708.7 keV stretched quadrupole (\textit{E}2) transition.}      &  &3.0 & \textit{M}1+\textit{E}2\\

&403.1& 0.80(0.08) & 3$^{+}$   & 2$^{+}$    & 0.56(0.08)$^d$     & &  &\textit{M}1+\textit{E}2 \\

&1086.1& 4.11(0.42) & 3$^{+}$   & 2$^{+}$    & 0.82(0.10)$^d$     & $+$0.01(0.04)  &0.29 & \textit{M}1+\textit{E}2\\

2026.9 &543.3& 1.56(0.17)  & 4$^{+}$   & 4$^{+}$    & 0.83(0.11)$^d$     & $-$0.11        (0.09)& 2.2  & \textit{M}1+\textit{E}2\\

& 635.5& 7.17(0.72) & 4$^{+}$   & 2$^{+}$    & 1.00(0.11)\footnote{DCO ratios are obtained from 618.1 keV stretched quadrupole (\textit{E}2) transition.}     & $+$0.14       (0.02) & & \textit{E}2\\

&1318.3& 0.85(0.10) & 4$^{+}$   & 2$^{+}$    & 0.99(0.16)$^d$     &   & &\textit{E}2\\

2217.2 &733.7& 74.24(7.56) & 6$^{+}$   & 4$^{+}$    & 0.98(0.11)$^d$     & $+$0.12(0.01)  & &  \textit{E}2\\

2450.1 & 655.3& 3.93(0.40) & 5$^{+}$   & 3$^{+}$    & 0.97(0.11)$^d$     & $+$0.05( 0.04)  & & \textit{E}2\\

&967.0& 0.56(0.10) & 5$^{+}$   & 4$^{+}$    & 0.58(0.12)$^d$    &   &$-$3.8 & \textit{M}1+\textit{E}2\\

2645.1 &618.1& 4.11(0.42) & 6$^{+}$   & 4$^{+}$    & 0.98(0.11)$^d$     & $+$0.23(0.07)  & & \textit{E}2\\

&1162.1& 0.57(0.07) & 6$^{+}$   & 4$^{+}$    & 0.96(0.16)$^d$    & $+$0.44(0.21)  & & \textit{E}2\\

2694.7 &1210.7& 0.43(0.1) & (4$^{+}$)   & 4$^{+}$    &      &   & &(\textit{M}1+\textit{E}2)\\

&1303.7& 0.25(0.04) & (4$^{+}$)   & 2$^{+}$    & 0.91(0.27)$^d$     &   & &(\textit{E}2)\\

3084.7 & 634.6& 2.14(0.26) & 7$^{+}$   & 5$^{+}$    & 0.88(0.12)\footnote{DCO ratios are obtained from 774.8 keV stretched quadrupole (\textit{E}2) transition.}     & $+$0.06       (0.03)  & & \textit{E}2\\

3088.4&871.2& 33.52(3.41) & 8$^{+}$   & 6$^{+}$    & 0.98(0.11)$^d$     & $+$0.12(0.01)  & & \textit{E}2\\

3504.5 &859.5& 0.27(0.04) & 8$^{+}$   & 6$^{+}$    &  0.93(0.14)$^e$    &  $+$0.07(0.06) &  &\textit{E}2\\

&1287.0& 0.40(0.15) & 8$^{+}$   & 6$^{+}$    &      &   & &(\textit{E}2)\\

3909.7 &825.0& 0.95(0.11) & 9$^{+}$   & 7$^{+}$    & 0.91(0.14)$^d$     &  $+$0.11(0.06) & & \textit{E}2\\

\end{tabular}
\end{ruledtabular}
\end{table*}

\subsection{Band I}
Band-I, earlier reported up to $ I^{\pi} = 8^{+} $ at 3089 keV \cite{Mol,114te2,114te3}, is a positive parity band built on the ground state configuration. All the previously reported transitions of this band have been identified in the present work. Spin and parity of the states have been adopted from Ref.~\cite{Mol,114te2} after verification from the present spectroscopic results. 
\subsection{Band II}
A sequence of two $ \gamma $-rays marked as band 5 in Ref.~\cite{114te2}, was reported up to E$ _{(6^{+})} = 2644 $ keV with tentative spin and parity assignment. This sequence, which is found to decay to the ground state band via 682.3 ($ \Delta I = 0 $) and 1391.2 ($ \Delta I = 2 $) keV $ \gamma $-rays, has been extended further up to 
E$ _{8^{+}} = 3504.5 $ keV by placing a 859.5 keV $ \gamma $-ray. Three new $ \Delta I = 2 $ $ \gamma $ transitions, \textit{viz.}, 1318.3 ($ 4^{+}_{2} \rightarrow 2^{+}_{1} $), 1162.1 ($ 6^{+}_{2} \rightarrow 4^{+}_{1} $) and 1287.0 ($ 8^{+}_{2} \rightarrow 6^{+}_{1} $) keV have been placed in the level scheme. Energy gated spectrum of 708.7 keV has been presented in Figs.~\ref{gate-709} in favour of present placement. Spin and parity of all the states belonging to this band have been assigned on the basis of present spectroscopic results. Both the $\Delta I = 0$ transitions, \textit{viz.}, 543.3 and 682.3 keV, were found to have significant \textit{E}2 admixture as estimated from the mixing ratio extracted from the present DCO ratio (Table~\ref{table1}).
\subsection{Band III}
A state at 1794.7 keV, decaying mainly to the $ I^{\pi} = 2^{+} $ state at 708.7 keV via 1086.1 keV $\gamma$-transition, was reported with a tentative spin/parity $ I^{\pi} = (2^{+}) $ assignment from the $ \beta^{+} $ decay study of the $^{114}$I \cite{zimmerman,NDS-114Te}. Apart from this strongest decay branch, this state was reported to decay via another three branches via 310.7, 403.0 and 1793.4 keV $\gamma$-transitions \cite{zimmerman,NDS-114Te}. Three (\textit{viz.} 310.9, 403.1 and 1086.1 keV) out of these four decay paths have also been identified in the present work (\figurename~\ref{gate-655}). The experimental DCO ratio of these three transitions supports present $ I^{\pi} = 3^{+} $ assignment for this state. According to the present spin/parity assignment, the multipolarity of the 1795 keV transition becomes $ \Delta I = 3 $, which may be one of the causes for non-observation of this transition in this study. The available branching ratio of the 1795 keV state also indicates a very low intensity ($ I_{\gamma} \approx 0.4 $) of this transition, which is again beyond the present detection limit. A sequence of three $\gamma$-transitions, \textit{viz.} 655.3, 634.6 and 825.0 keV (marked as Band-III), has been established above this 1794.7 keV state on the basis of $ \gamma \gamma $ coincidence study. The 708.7 and 655.3 keV energy gated spectra have been shown in favour of the present placement (\figurename~\ref{gate-709}, \ref{gate-655}). Spin and parity of all the states belonging to this sequence have been assigned firmly on the basis of angular correlation and linear polarization results, as listed in \tablename~\ref{table1}.
\begin{figure}[h]
\begin{center}
\includegraphics[width=\columnwidth]{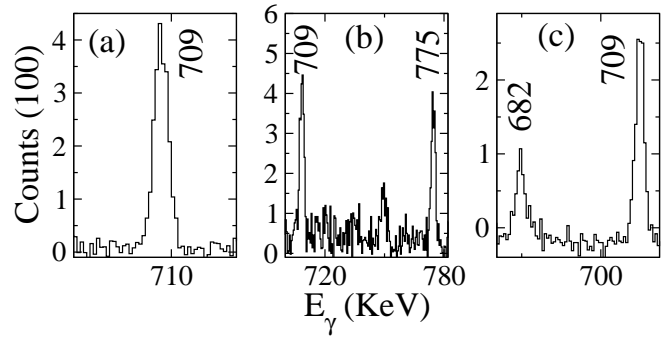}
\caption{Prompt $ \gamma \gamma $ coincidence spectra gated by (a) 639.5 keV, (b) 1210.7 keV, and (c) 1303.7 keV transitions.}
\label{gate-640_1211_1304}
\end{center}
\end{figure}
\subsection{Other non-yrast states}
Apart from the above mentioned groups of states (Bands I-III), two more excited states at 1348.2 and 2694.7 keV have also been observed in this work related to the present interest. These two states were reported earlier in the $ \beta^{+} $ decay study of the $^{114}$I \cite{zimmerman}. The 1348.2 (2694.7) keV state with $ I^{\pi} = (0^{+}) $ ($ I^{\pi} = (4^{+}) $) was found to decay to the 708.7 (1483.5/1391.2) keV state(s) via 639.5 (1210.7/1303.7) keV  $\gamma$-transition(s). Energy gated spectra of 639.5, 1210.7 and 1303.7 keV $\gamma$-transitions have been shown in \figurename~\ref{gate-640_1211_1304}, to confirm these placements. The DCO ratios of 639.5 and 1303.7 keV $\gamma$-transitions are indicating the quadrupole nature of these transitions.
\section{Discussion}
\begin{figure}[!]
\begin{center}
\includegraphics*[scale=0.31, angle = 0]{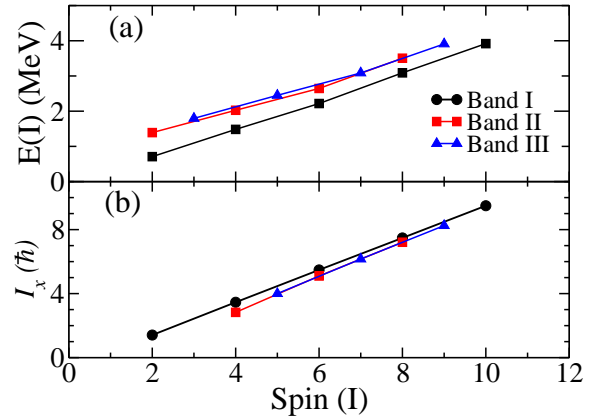}
\caption{(a) The experimental excitation energies of the band I, band-II and band-III, respectively with spin. (b) Projection of total angular momentum on the rotation axis (I$_x$) of the band-I, band-II and band-III plotted with spin.}
\label{Ivsw} 
\end{center}
\end{figure}

\begin{figure}[t]
\begin{center}
\includegraphics*[scale=0.35, angle = 0]{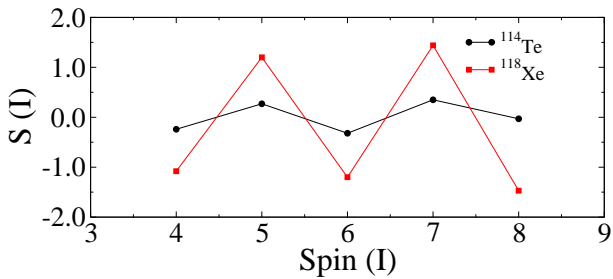}
\caption{Calculated odd-even energy staggering S(I) plotted against spin (I) for $^{114}$Te. 
Values for $^{118}$Xe are taken from literature~\cite{xe1,xe2}.}
\label{SI}
\end{center}
\end{figure}

\begin{figure*}[t]
\begin{center}
\includegraphics*[width=10.0cm, angle = -90]{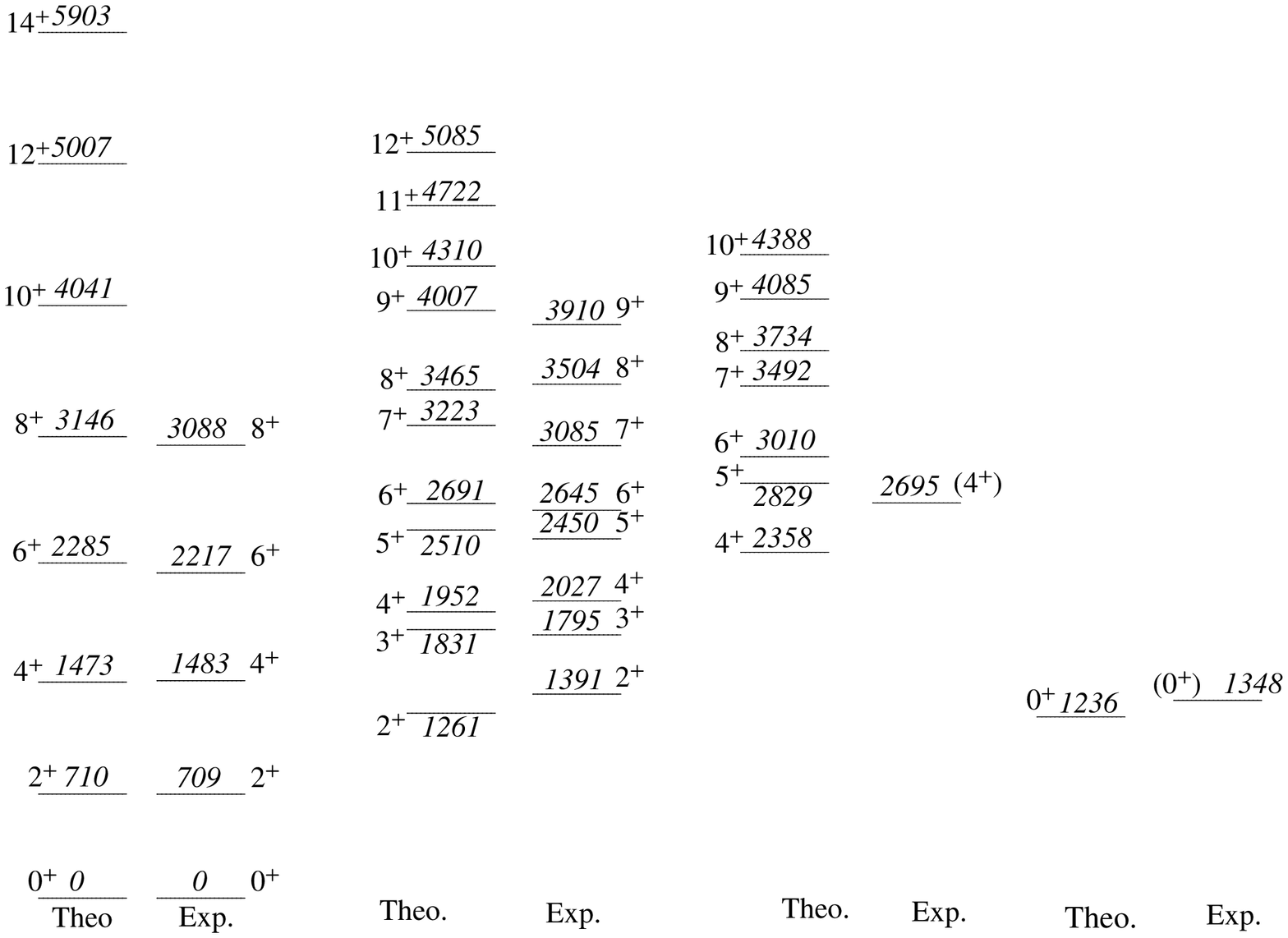}
\caption{Comparison of experimental and the calculated band energies using IBM formalism in $^{114}$Te. The level energy associated with the states are given in keV.}
\label{ibm}
\end{center}
\end{figure*}

\begin{figure}[t]
\begin{center}
\includegraphics*[width=6.5cm, angle = 0]{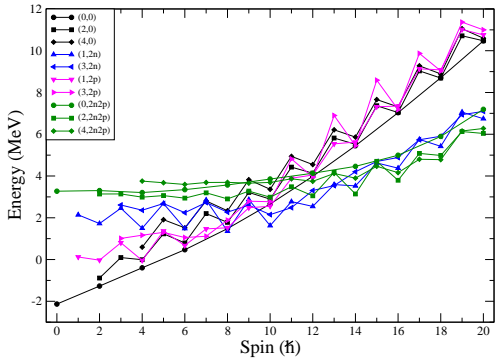}
\caption{Theoretical band diagram for $^{114}$Te. The labels $(K,\#)$
characterize the states, with $K$ denoting the $K$ quantum number
and $\#$ the number of quasiparticles. For example, (0,0), (2,0),
and (4,0) correspond to the $K=0$ ground-, $K=2$ $\gamma$-, and
$K=4$ $\gamma$ $\gamma$-band, respectively, projected from the 0-qp
state. (1,2n), (3,2n), (1,2p), (3,2p), (2,4), and (4,4) correspond,
respectively, to the projected 2-neutron-aligned state,
2-proton-aligned state, 2-neutron-plus-2-proton aligned state, with
different $K$ quantum numbers.}
\label{tpsm1}
\end{center}
\end{figure}

Low-lying structure of $^{114}$Te mainly consists of three sequences of \textit{E}2 transitions. Apart from the strongly populated ground state band (I), two more band structures were established in the present work. The excitation energy and the projection of the total angular momentum on the rotation axis ($I_x$) of these bands are plotted with spin, as shown in Figs.~\ref{Ivsw}(a) and \ref{Ivsw}(b). The excitation energy and $I_x$ versus spin plots show similar slopes for the ground-state band (Band I) and the other two bands (Band II and Band III), indicates that the dynamic moment of inertia of these bands is nearly equal. Coupling of a phonon excitation with a rotational configuration can give rise to similar moment of inertia \cite{106Mo,sc}. Therefore, similar characteristics for all the three bands are indicating that the band-II and band-III correspond to the quasi-$\gamma$-vibrational structure. The properties of $ \Delta I = 0 $ transitions, as pointed out in Refs.~\cite{saikat,kusakari}, are also found consistent in the present case. The \textit{E}2 fraction of $ 2^{+}_{1} \rightarrow 2^{+}_{2} $ and $ 4^{+}_{1} \rightarrow 4^{+}_{2} $ $ \Delta I = 0 $ transitions are found decreasing with increasing spin. The odd-even energy staggering [S(I)]~\cite{McCutchan,Casten-1} plotted as a function of spin also shows a similar pattern when compared with the quasi-$\gamma$ band of $^{118}$Xe (as shown in Fig.~\ref{SI}) with minimum at even spin indicating the band structure observed in $^{114}$Te is of quasi-$\gamma$ vibrational in nature. However, the absolute values of staggering are small in the present case compared to $^{118}$Xe~\cite{McCutchan,xe1,xe2}. 
The calculated value of E$_s$ /E(2$^+_1$ ) is -0.13 which indicates a $\gamma$-soft nature for the nucleus~\cite{HW}.

The experimental results have been discussed below in the light of both Interacting Boson Model (IBM) and Triaxial Projected Shell Model (TPSM) calculations.


\subsection{Interacting Boson Model approximation}




\begin{figure*}[t]
\begin{center}
\includegraphics*[width=10.0cm, angle = -90]{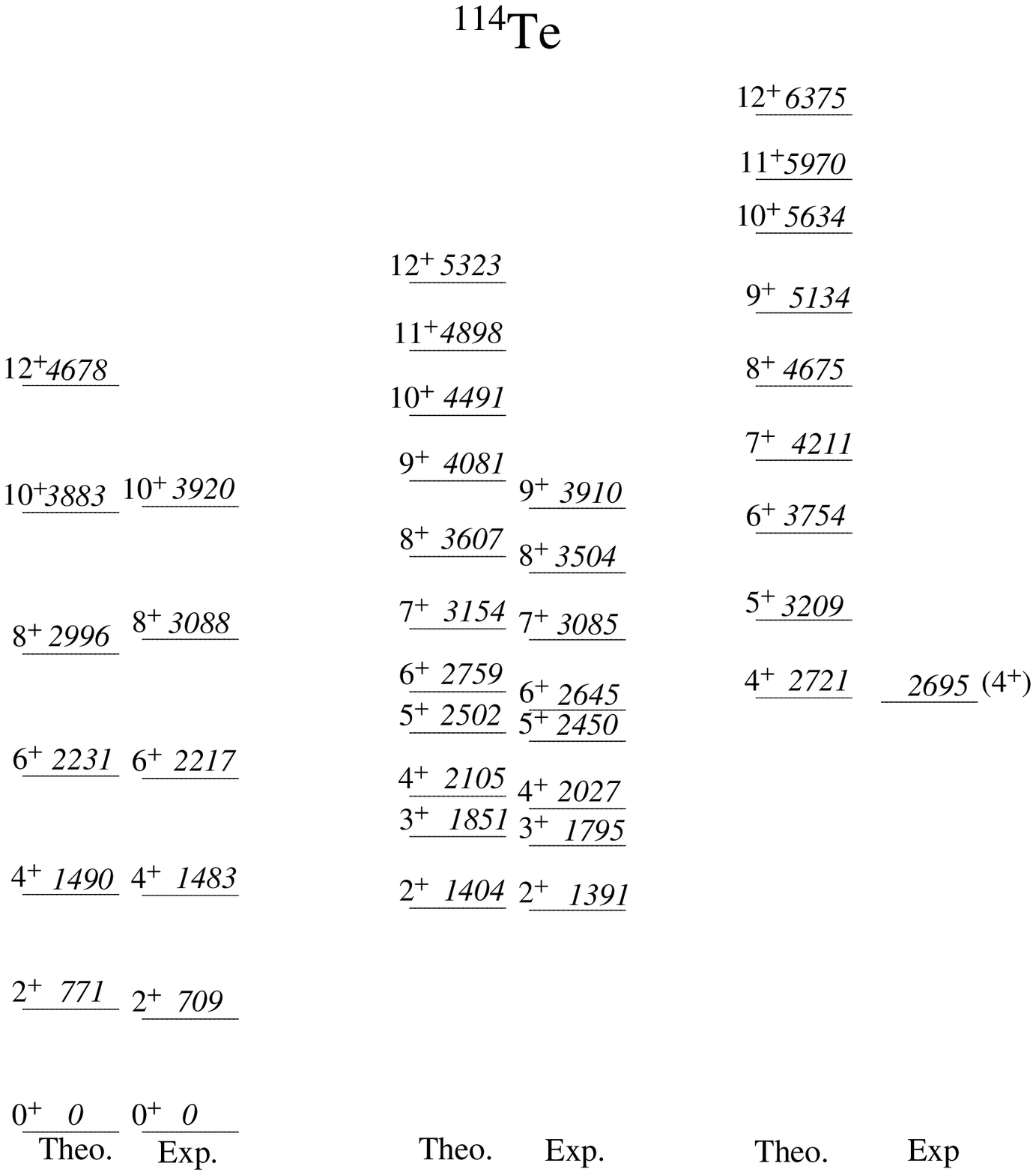}
\caption{Comparison of experimental and the calculated band energies using TPSM formalism in $^{114}$Te. The level energies associated with the states are given in keV.}
\label{tpsm2}
\end{center}
\end{figure*}




Nuclei in the neighbourhood of $Z=50$ shell closure are known to be anharmonic
vibrators and hence, are expected to be close to the 
U(5) limit of the interacting boson model (IBM). For example 
Van Ruyven {\em et al.}~\cite{118te} found that excited states in $^{118,120}$Te  can be very well 
understood in terms of IBM. 
Ground bands of Sn nuclei have been described by U(5) symmetry~\cite{Singh}. 

However, there is evidence that significant departure from the U(5) symmetry is
also possible in some of the nuclei in this region. 
Cd nuclei with 
$Z=48$ are can be described by a U(5) as well as a non-U(5) Hamiltonian~\cite{Long}.   
M\"{o}ller {\em et al.}~\cite{Mol} have studied the excited states and the
electromagnetic transition probabilities in $^{114}$Te. They concluded that 
though the ground band can be explained well in terms of the U(5) limit of 
IBM, it fails to explain the trend of the B(E2) values.  

Here, the yrast band of $^{114}$Te is known up to the 8$^+$ state. Though this 
band can be easily described by the U(5) model, we found that the 
quasi-$\gamma$ vibrational band
predicted by this model shows strong staggering in disagreement with experimental data. For this reason, we decided to use 
a more general IBM-1 Hamiltonian. Since the number of bosons is not too large 
and the ground band is easily described by the U(5) limit of the Hamiltonian, 
we considered a Hamiltonian which has an additional quadrupole-quadrupole interaction
consistent with the O(6) limit. 
A similar Hamiltonian was used by McCutchan {\em et al}~\cite{McCutchan} in their study
on Ba, Xe and Ce isotopes.  
The two parameter Hamiltonian they used was
\begin{equation} H=C\left[(1-\zeta)n_d-\frac{\zeta}{4N_B}Q^\chi\cdot Q^\chi\right]\end{equation}
where 
\begin{eqnarray} n_d=(d^\dagger \tilde{d})_0\nonumber\\
Q^\chi=(d^\dagger\tilde{s})_2+(s^\dagger\tilde{d})_2-\chi
(d^\dagger \tilde{d})_2
\end{eqnarray}
written in terms of the $d$ boson
creation and annihilation operators~\cite{Bon}. The two parameters are $C$ and $\zeta$
and $N_B$ is the number of bosons.  the parameter $\chi$ was not taken as free. 
In the U(5) limit, the second term is absent. In the O(6) limit, $\chi=0$ and 
the first term does not occur.
Saxena {\em et al.} have also used a similar Hamiltonian for IBM-2~\cite{ms} to 
describe heavier Te isotopes where they have varied the value of $\chi$.
In accordance with O(6) symmetry,  we chose $\chi$ as zero and treated the 
coefficients of the other two terms as free parameters. 
We fitted the excitation energies by  minimizing the sum of errors in energy prediction to extract the 
parameters. However, our results were not very satisfactory. 
We then decided to extend our Hamiltonian by adding another term,   
\begin{equation}
L=\sqrt{10}(d^\dagger \tilde{d})_1\end{equation}
so that our Hamiltonian can be written as
\begin{equation}
H=\epsilon n_d+a_1(L\cdot L)+a_2(Q\cdot Q)\end{equation}

  
The parameters obtained after fitting are $\epsilon=0.497$ MeV, 
$a_1=0.02548$ MeV, and $a_2=0.042$ MeV.
The rms deviation for fourteen excited states is 117.63 keV.
The result of the three parameter fitting is shown in Figure \ref{ibm}. As
one can see, the agreement is reasonably good.
The coefficient of the quadrupole-quadrupole interaction term is of typical
magnitude and has a significant effect on the staggering in the
quasi-$\gamma$ vibrational band. Hence, we may conclude that this nucleus
falls between the U(5) and the O(6) symmetries. Furthermore, 
we see that the ratio of the excitation energy values of the first $4^+$ state to the 
first 2$^+$ state has the value 2.09. 
Similarly, the ratio of the energy of the quasi-$\gamma$ bandhead
to that of the first $2^+$ state is 1.96. 
These values, along with the results of the calculation, indicate that the
nucleus $^{114}$Te lies between the vibrator and the E(5) critical point of the Casten
symmetry triangle \cite{Casten} on the arm connecting the U(5) and the O(6) dynamical symmetries.

\subsection{Triaxial Projected Shell Model Calculation}
Multi-quasiparticle TPSM approach has been
developed and it has been shown to provide a consistent
description of yrast, $\gamma$ $(K = 2)$ and $\gamma\gamma$ $ (K = 4)$ bands
in transitional nuclei \cite{JK99,SJ00,sc}. In this method, the three-dimensional projection technique is employed to project out
the good angular-momentum states from product states built
upon quasiparticle  configurations of triaxially deformed
Nilsson+BCS model. The shell model Hamiltonian is then
diagonalized in this angular-momentum projected basis. The
TPSM space includes multi-quasiparticle states; hence, is
capable of describing near-yrast band structures at high-spins.

The TPSM basis employed in this study consists of
0-quasiparticle vacuum, two-proton, two-neutron, and four-quasiparticle
configurations \cite{GJ08}. The quasiparticle basis chosen is adequate to
describe high-spin states up to angular momentum I $\sim$ 20. In
the present analysis, we shall, therefore, restrict our discussion
with this spin regime.

As in the earlier TPSM calculations, we use the pairing plus 
quadrupole-quadrupole Hamiltonian \cite{KY95}
\begin{equation}
\hat H = \hat H_0 - {1 \over 2} \chi \sum_\mu \hat Q^\dagger_\mu
\hat Q^{}_\mu - G_M \hat P^\dagger \hat P - G_Q \sum_\mu \hat
P^\dagger_\mu\hat P^{}_\mu,
\label{hamham}
\end{equation}
where $\hat H_0$ is the spherical single-particle Hamiltonian, which
contains a proper spin-orbit force \cite{Ni69}.
$\chi$ is the strength of the quadrupole-quadrupole force related
in a self-consistent way to deformation of the quasiparticle
basis and $G_M$ and $G_Q$
are the strengths of the monopole and quadrupole
pairing terms, respectively. The configuration space employed
corresponds to three principal oscillator shells $\nu[3,4,5]$  and
$\pi[2,3,4]$. The pairing strengths have been parametrized
 in terms of two constants $G_1$ and $G_2$. In
this work, we choose $G_1$ = 21.14 MeV and $G_2$ = 13.86 MeV;
with these pairing strengths we approximately reproduce the
experimental odd-even mass differences in this region.
 The quadrupole pairing strength $G_Q$ is assumed to be proportional
to $G_M$, and the proportionality constant was set to 0.18. 
 These interaction strengths are
consistent with those used earlier for the same mass region
\cite{GH14,bh15,JG09}.

\begin{figure}[t]
\begin{center}
\includegraphics*[width=7.0cm, angle = -90]{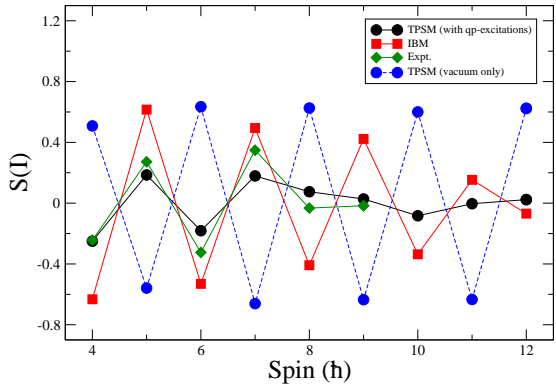}
\caption{Comparison of observed, TPSM and IBM calculated staggering
  parameter, Eq. (\ref{SI}), for the quasi-$\gamma$ band in $^{114}$Te.
  TPSM solid curve (Black color) is  with quasiparticle excitations and dotted curve (Blue color) is  with vacuum only}
\label{tpsm3}
\end{center}
\end{figure}

TPSM calculations proceed in several stages. In the first stage, the
deformed basis space is constructed by solving the
triaxially-deformed Nilsson potential. In the present work,
the axial deformation parameter
$\epsilon=0.150$  adopted from the Ref. \cite{PM95}.
The non-axial deformation parameter $\epsilon' = 0.09$
has been chosen so that the behavior of the $\gamma$
band is properly described.

In the second step, the good angular-momentum states are obtained from the deformed basis by employing the three-dimensional angular-momentum projection technique. The projected bands obtained from 0-, 2-, and 4-quasiparticle states close to the Fermi surface are displayed in Fig.~\ref{tpsm1} so-called band diagram, which are the diagonal matrix elements before band mixing. The projection from the 0-quasiparticle configuration gives rise to band structures with $K=0,2,4$, corresponding to the ground-, $\gamma$- and $\gamma \gamma$-band. The calculated band-head energies of the $\gamma$- and $\gamma\gamma$-bands are about 1.2452 MeV and 2.7252 MeV, respectively, above the ground state.

\begin{figure}[t]
\begin{center}
\includegraphics*[width=7.0cm, angle = 0]{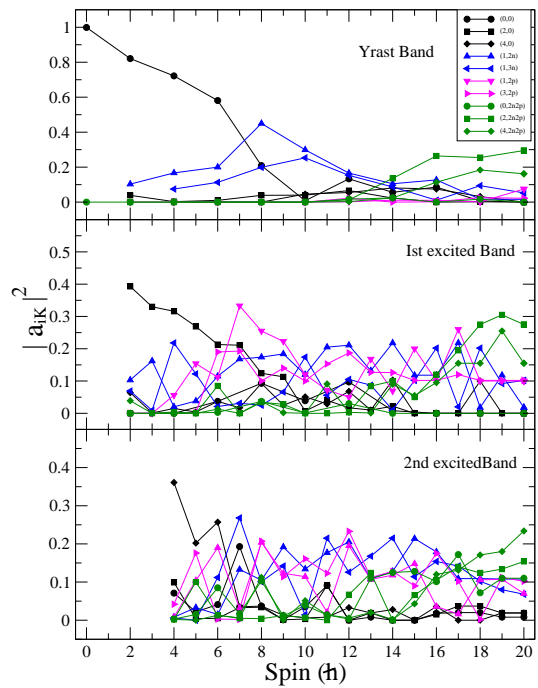}
\caption{Probabilities of the projected configurations in the yrast-, 1st-,
and 2nd-excited bands.}
\label{tpsm4}
\end{center}
\end{figure}

In the third and the final stage, the projected basis are used to diagonalize the shell model Hamiltonian, Eq.~(\ref{hamham}). The band energies, obtained after diagonalization, are shown in Fig.~\ref{tpsm2} with the available experimental data. It is evident from the figure that TPSM results are in excellent agreement with the known experimental energies. In Fig.~\ref{tpsm2}, the calculations slightly overestimate the bandhead energy of the $\gamma\gamma$-band, and we hope that this well-developed band will be populated in future experimental studies. The first quasiparticle ${(h_{ 11/2}}^2)$ two neutron alignment is predicted around I = 8, and the transition to a four quasi-particle $\nu{(h_{11/2})}^2$ $\pi {(g_{9/2})}^2$ band is expected to occur around \textit{I} = 14.

In order to understand the  importance of the $\gamma$ degree of freedom
in the description of the triaxial shape in
$^{114}$Te, the staggering parameter, defined as,
\begin{equation}
S(I) = \frac{[E(I)-E(I-1)]-[E(I-1)-E(I-2)]}{E(2^{+}_1)} \label{SI}
\end{equation}
is plotted for the quasi-$\gamma$ band in Fig.~\ref{tpsm3}
with vacuum configuration only and with all the
quasiparticle configurations. In the same
figure, we also provide results of the IBM approach which in good agreement with experimental data.
It is evident from
Fig.~\ref{tpsm3} that the phase of the staggering of the quasi-$\gamma$ band
with vacuum configuration only and with the inclusion of the
quasiparticle excitations is opposite to each other. The phase with the 
vacuum configuration only  has odd-spin
states lower than the even-spin states and corresponds to
Davydov-Filippov or $\gamma$-rigid motion \cite{AS58}. The inclusion of the
quasiparticle excitations is shown to reverse the staggering phase with
even-spin states lower than the odd-spin states as in the limiting
case of Wilet-Jean or $\gamma$-soft motion \cite{lw}. 
Therefore, the inclusion of the quasiparticle excitations transforms the
motion from $\gamma$-rigid to $\gamma$-soft.
The phase and the
magnitude of the staggering after the inclusion of the quasiparticle 
excitations is in good agreement with the
corresponding experimental numbers and, therefore, it can state that the 
$^{114}$Te nucleus is $\gamma$-soft.
The TPSM results further indicate that the above spin $I=8$, the staggering
amplitudes become smaller, and the reason for this is due to a
considerable mixing of the two-quasiparticle configurations with the
quasi-$\gamma$ band at higher spins. In order to probe the mixing, the
probabilities of various projected configurations are plotted in
Fig.~\ref{tpsm4} for the yrast, the 1st-, and the 2nd-excited bands.
The yrast band up to $I=6$ is dominated by the 0-quasiparticle configuration
with (0,0) i.e., $K=0$, and above this spin the 2-neutron-aligned band $[(1,2n)]$ is the
dominant configuration. Above $I=14$, the yrast band is primarily
composed of 4-quasiparticle configurations $[(2,2n2p)]$. The 1st-excited band has the
dominant  $K=2$ 0-quasiparticle configuration $[(2,0)]$ until $I=6$ and, therefore, is the
$\gamma$-band. However, above $I=6$, the 1st-excited band has  two-quasiparticle dominant component $[(1,2p)]$. 
The 2nd excited band has a dominant  $K=4$ 0-quasiparticle
configuration $[(4,0)]$, referred to as $\gamma\gamma$-band, up to $I=6$.
Above this spin value, mixed structures are obtained. The (1,2n), (3,2n), (1,2p) and (3,2p)
state from the two-quasiparticle configuration seems to become important along
with some four-quasiparticle configurations.

\section{Summary}

Excited states of $^{114}$Te were populated by means of light-ion induced fusion evaporation reaction $^{112}$Sn($^{4}$He, 2n)$^{114}$Te at 37 MeV beam energy. Several new $ \gamma $-transitions were placed in the level scheme of $^{114}$Te, based on the $ \gamma$ - $ \gamma $ coincidence and relative intensity measurements. The spin and parity of the levels were assigned on the basis of angular correlation and polarisation asymmetry measurements. Theoretical calculations under the framework of the Triaxial Projected Shell Model and Interacting Boson Model were performed in order to interpret the experimental results. Both the model calculations were found in well-agreement with the experimental results, thereby, confirming the existence of the quasi-$\gamma$ band structure in the $^{114}$Te nucleus. Further, systematic experimental investigation of low-lying states are required in order to understand the shape evolution in Te isotopes.

\section*{Acknowledgement}

The authors gratefully acknowledge the effort of the cyclotron operators for providing a good quality of $^{4}$He beam. We would like to thank all the members of INGA collaboration for their support and active involvement. 
INGA is partially funded by Department of Science and Technology, Government of India (No. IR/S2/PF-03/2003-II).
S. R. would like to acknowledge the financial assistance from the University Grants Commission - Minor Research Project [No. PSW-249/15-16 (ERO)]. G. G. acknowledges the support provided by the University Grants Commission - departmental research support (UGC-DRS) program. P. R. acknowledges the support of the UGC-NET fellowship (Sr. No. 2121551170). H. P. is grateful for the support of the Ramanujan Fellowship research grant under SERB-DST (SB/S2/RJN-031/2016).
H. P. is also grateful for the fruitful discussion with Prof. Sukalyan Chattopadhyay (SINP, Kolkata).

\end{document}